\documentclass[journal]{IEEEtran}
\usepackage{cite}

\usepackage{ulem}

\ifCLASSINFOpdf
   \usepackage[pdftex]{graphicx}
    \DeclareGraphicsExtensions{.pdf,.jpeg,.png,.eps}
\else
  \usepackage[dvips]{graphicx}
  \DeclareGraphicsExtensions{.eps, .pdf}
\fi
\usepackage{amsmath,amssymb}
\usepackage{algorithmic}
\ifCLASSOPTIONcompsoc
  \usepackage[caption=false,font=normalsize,labelfont=sf,textfont=sf]{subfig}
\else
  \usepackage[caption=false,font=footnotesize]{subfig}
\fi

\ifCLASSOPTIONcaptionsoff
  \usepackage[nomarkers]{endfloat}
 \let\MYoriglatexcaption\caption
 \renewcommand{\caption}[2][\relax]{\MYoriglatexcaption[#2]{#2}}
\fi

\usepackage{color}
\usepackage{bm}

\begin{document}
%

\title{Neural Turbo Equalization: Deep Learning \\for Fiber-Optic Nonlinearity Compensation}

%
%
%

\author{Toshiaki~Koike-Akino,~\IEEEmembership{Senior~Member,~IEEE,~Senior~Member,~OSA},
Ye~Wang,~\IEEEmembership{Senior~Member,~IEEE},
 David~S.~Millar,~\IEEEmembership{Member,~IEEE,~Member,~OSA},
 Keisuke~Kojima,~\IEEEmembership{Senior~Member,~IEEE,~Fellow,~OSA},
 Kieran~Parsons,~\IEEEmembership{Senior~Member,~IEEE,~Member,~OSA},
\thanks{The authors are with
Mitsubishi Electric Research Laboratories (MERL), 201 Broadway,
Cambridge, MA 02139, USA (e-mail: koike@merl.com; yewang@merl.com; 
millar@merl.com; kojima@merl.com; parsons@merl.com).}
\thanks{This paper contains in part our previous work\cite{Koike-SPPCom14, Koike-SPPCom18, Koike-ECOC19}.}
}

\maketitle

\begin{abstract}
 Recently, data-driven approaches motivated by modern deep learning 
 have been applied to optical communications in place of traditional model-based counterparts.
 The application of deep neural networks (DNN) allows flexible statistical
 analysis of complicated fiber-optic systems without relying on any specific physical models.
 Due to the inherent nonlinearity in DNN, various equalizers based on DNN have shown significant potentials to mitigate fiber nonlinearity.
 In this paper, we propose a turbo equalization (TEQ) based on DNN as a new alternative framework to deal with nonlinear fiber impairments for future coherent optical communications.
 The proposed DNN-TEQ is constructed with nested deep residual networks (ResNet) to train extrinsic likelihood given soft-information feedback from channel decoding.
 Through extrinsic information transfer (EXIT) analysis, we verify that our DNN-TEQ can accelerate decoding convergence to achieve a significant gain in achievable throughput by 0.61\,b/s/Hz.
 We also demonstrate that optimizing irregular low-density parity-check (LDPC) codes to match EXIT chart of the DNN-TEQ can improve achievable rates by up to 0.12\,b/s/Hz.
\end{abstract}


%
\IEEEpeerreviewmaketitle

\section{Introduction}
%
%
%
%

\IEEEPARstart{M}{achine} learning techniques\cite{Hinton-DBN, LeCun, Hockreiter-LSTM} have been recently applied to optical communications systems
to deal with various issues such as network monitoring\cite{Tanimura2016, Khan2017, Wang2017}, traffic control\cite{Guo2018, Yu2018, Luo-RL, Tang-RL}, signal design\cite{Ye2018, Karanov2018, Li2018e2e, Jones2018, Chagnon2018}, and nonlinearity compensation\cite{RiosMuller, Chuang2017, Kamalov2018, Li2018, Koike-SPPCom18, Koike-ECOC19}.
Since the fiber nonlinearity is a major limiting factor to the achievable information rates\cite{Ellis-JLT-2009, Secondini-JLT-2013, Renaudier-2007},
mitigating nonlinearity has been of great importance to realize high-speed, reliable, and long-reach optical communications.
Conventionally, a number of model-based nonlinear equalizers to compensate for fiber distortion were investigated, e.g.,
maximum-likelihood sequence equalizer (MLSE)\cite{Alic, Cai, Koike-SSE},
turbo equalizer (TEQ)~\cite{Djordjevic-TEQ, Batshon-TEQ, Duan-TEQ},
Volterra series transfer function (VSTF)~\cite{Guiomar-2013, Guiomar-ECOC-2011},
and digital backpropagation (DBP)~\cite{Ip, Irukulapati-SDBP-2013, Yan-PDBP-2011, Ip-OFC-2011}.
However, those nonlinear equalizations are computationally complex and susceptive to model parameter mismatch in general.
Recent data-driven approaches motivated by deep learning can favorably replace such traditional model-based methods as the use of deep neural networks (DNN) allows flexible statistical analysis of complicated fiber-optic systems without relying on specific models.
In the past few years, DNN has shown its high potential in nonlinear performance improvement, e.g., \cite{RiosMuller, Chuang2017, Kamalov2018, Li2018, Koike-SPPCom18, Koike-ECOC19, Karanov2018, Li2018e2e, Jones2018, Chagnon2018}.

Nonetheless, most existing work did not appropriately account for practical interaction with forward error correction (FEC) codes.
For example, multi-class soft-max cross-entropy loss is often used to train DNN, which is relevant only when nonbinary FEC codes are assumed.
For more practical bit-interleaved coded modulation (BICM) systems, it was found in~\cite{Koike-SPPCom18} that binary cross-entropy (BCE) loss can improve accuracy and scalability to high-order quadrature-amplitude modulation (QAM).
In this paper, we propose a novel DNN application to perform TEQ for nonlinear mitigation in the context of BICM with iterative demodulation (ID).
Although DNN has already been popular in nonlinear compensation, our paper is the first attempt to adopt DNN for TEQ in the framework of BICM-ID which takes soft-decision feedback from the FEC decoder to refine the DNN output for improved equalization accuracy.
We make an analysis of the extrinsic information transfer (EXIT) of turbo DNN, and demonstrate that the proposed DNN paired with irregular low-density parity-check (LDPC) codes used in DVB-S2 standards offers a significant performance gain by accelerating the decoder convergence in nonlinear transmissions.

The contributions of this paper are summarized as follows:
\begin{itemize}
 \item {\bf{Trend overview}}: We first overview the recent trend of deep learning in optical society.
 \item {\bf{Multi-label DNN}}: We then verify that nonbinary cross-entropy is not scalable to high-order QAM signals and DNN trained with BCE loss can appropriately compensate for fiber nonlinearity.
 \item {\bf{Turbo DNN}}: We propose a nested residual DNN architecture for TEQ to further improve performance.
 \item {\bf{EXIT analysis}}: We analyze EXIT chart of our DNN-TEQ and show that DNN-TEQ accelerates decoding convergence.
 \item {\bf{LDPC design}}: We optimize degree distribution of LDPC codes to match EXIT charts of DNN-TEQ, achieving higher throughput.
\end{itemize}
Note that due to the above contributions, in particular the demonstration of rate improvement with optimized LDPC codes for DNN-TEQ, this paper is distinguished from our preliminary reports\cite{Koike-SPPCom14, Koike-SPPCom18, Koike-ECOC19}.
To the best of authors' knowledge, there is no other literature which applied DNN to TEQ for nonlinear compensation.

\section{Machine Learning for Optical Communications}

\subsection{Trend Overview}

Fiber-optic communications suffer from various linear and nonlinear
impairments, such as laser linewidth, amplified spontaneous emission
(ASE) noise, chromatic dispersion (CD), polarization mode dispersion
(PMD), self-phase modulation (SPM), cross-phase modulation (XPM),
four-wave mixing (FWM), and cross-polarization modulation
(XPolM)\cite{Ellis-JLT-2009, Secondini-JLT-2013, Renaudier-2007}. Although the
physics is well governed by nonlinear Schr\"{o}dinger equation (NLSE) model,
we may need high-complexity split-step Fourier method (SSFM) to solve lightwave propagation numerically.
It is hence natural to admit that the nonlinear physics necessitates nonlinear signal
processing to appropriately deal with the nonlinear distortions in practice.

In place of conventional model-based nonlinear signal processing,
the application of machine learning techniques\cite{Hinton-DBN, LeCun, Hockreiter-LSTM} to optical communication systems has recently received increased attention\cite{Tanimura2016, Khan2017, Wang2017, Guo2018, Yu2018,Luo-RL, Tang-RL, Ye2018, Karanov2018, Li2018e2e, Jones2018, Chagnon2018, RiosMuller, Chuang2017, Kamalov2018, Li2018, Koike-SPPCom18, Koike-ECOC19}. The promise of such data-driven approaches is that learning a black-box DNN could potentially overcome situations where limited models are inaccurate and complex theory is computationally intractable.

\begin{figure}[t]
 \centering
 \includegraphics[width=0.91\linewidth]{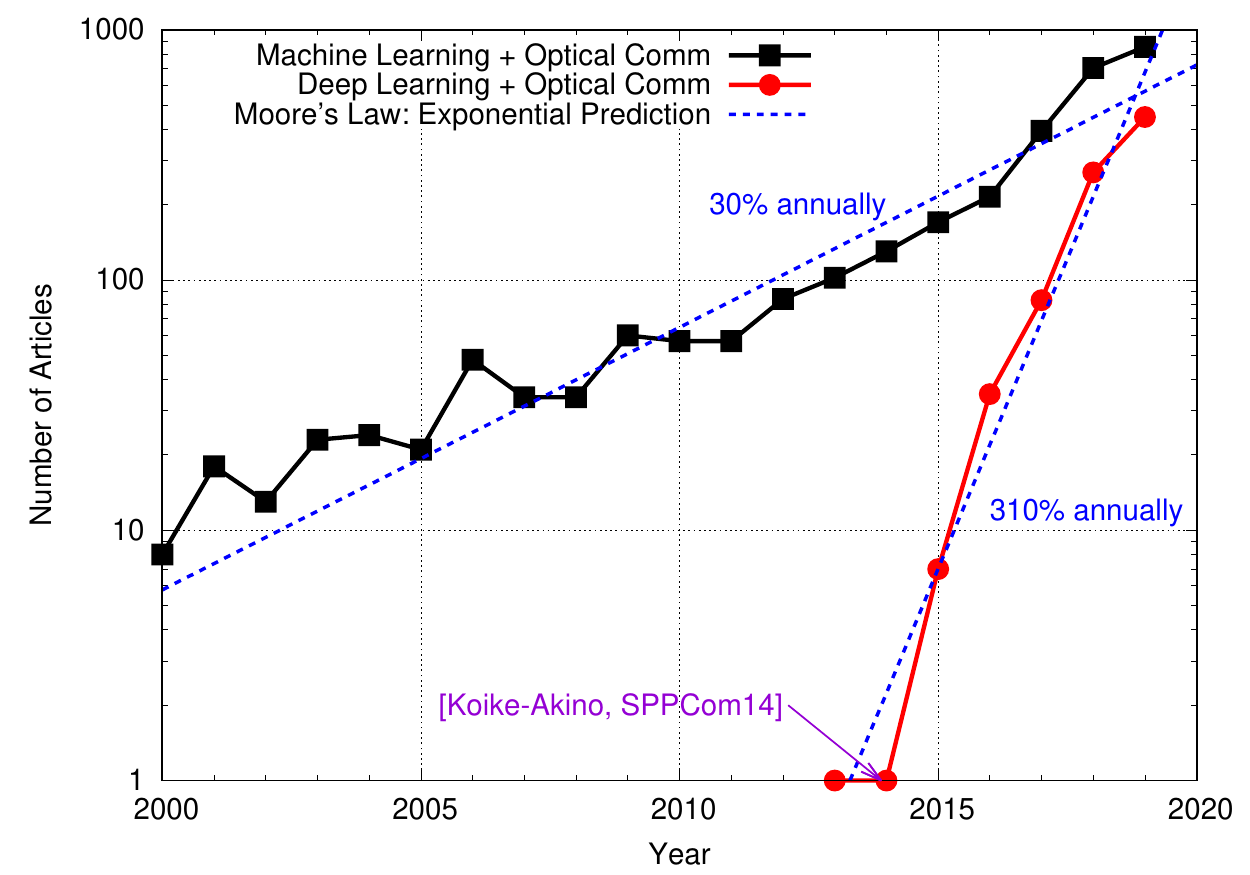}
 \caption{Machine/deep learning trend in optical communication applications (keyword hits on Google Scholar, excluding non-relevant ones).}
 \label{fig:trend}
\end{figure}

Fig.~\ref{fig:trend} shows the trend of machine learning applications in optical communications society in the past two decades.
Here, we plot the number of articles in each year according to Google Scholar search of the keyword combinations; ``machine learning'' + ``optical communication'' or ``deep learning'' + ``optical communication.''
As we can see, machine learning has been already used for optical communications since twenty years ago.
Interestingly, we discovered the Moore's law in which the number of applications exponentially grows by a factor of nearly $30\%$ per year.
For deep learning applications, more rapid annual increase by a factor of $310\%$ can be found in the past half decade.
As of today, there are nearly thousand articles of deep learning applications.
Note that the author's article\cite{Koike-SPPCom14} in 2014 is one of very first papers discussing the application of deep learning to optical communications.

\subsection{Statistical Learning Techniques}

\begin{figure}[t]
 \centering
 \includegraphics[width=.91\linewidth]{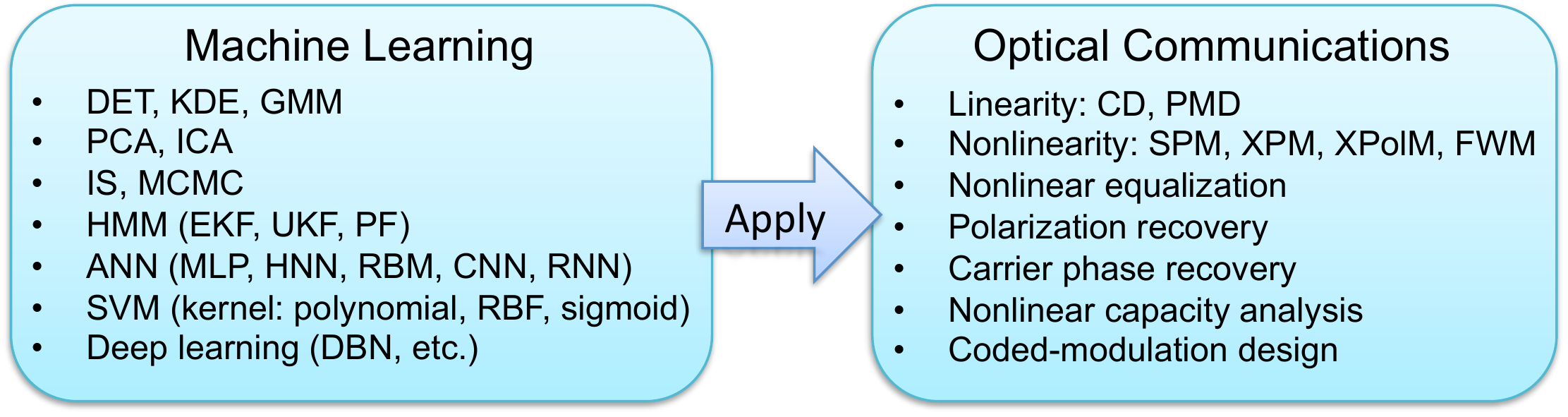}
 \caption{Machine learning approaches applied to optical communications\cite{Koike-SPPCom14}.}
 \label{fig:ml}
\end{figure}

We briefly overview some learning techniques to analyze nonlinear
statistics applied to optical communications as shown in Fig.~\ref{fig:ml}. For example,
density estimation trees (DET), kernel density estimation (KDE) and
Gaussian mixture model (GMM) can be alternative to
histogram analysis. Principal component analysis
(PCA) and independent component analysis (ICA) are useful to
analyze important factors of data.
For high-dimensional data sets, we may use Markov-chain Monte--Carlo
(MCMC) and importance sampling (IS). To analyze stochastic sequence data,
extended Kalman filter (EKF), unscented Kalman filter (UKF), and particle
filter (PF) based on hidden Markov model
(HMM) may be used.

Since mid-70's, artificial neural networks (ANN) have led
machine learning researches. Various topology including
multi-layer perceptron (MLP), Hopfield neural networks (HNN), restricted
Boltzmann machines (RBM), convolutional neural networks (CNN), and
recurrent neural networks (RNN) have been investigated. Since mid-90's,
support vector machine (SVM) has taken over the lead for machine
learning.
One of important techniques to analyze nonlinear statistics is kernel
trick, in which we analyze higher-dimensional linearlized feature
spaces called reproducing kernel Hilbert space (RKHS) with kernel functions including
radial basis function (RBF).
%
Since 2006, deep learning\cite{Hinton-DBN} based on DNN has been a major breakthrough in
media signal processing fields. In deep learning, many-layer deep belief networks (DBN) is trained with a
massively large amount of datasets. 

%

\subsection{Classic Machine Learning Applications}

Now, we show a few examples of machine learning approaches applied to
nonlinear fiber-optic communications. Xie {\it et al.} proposed the use
of ICA for polarization recovery\cite{Xie-ICA} as an alternative to
constant-modulus adaptation (CMA). Shallow ANN-based nonlinear equalizers have been studied in
literature\cite{Jarajreh, Giacoumidis2015, Zhao-NN}. We have
investigated GMM-based sliding MLSE and TEQ receivers\cite{Koike-SSE},
where up-to $2$\,dB performance improvement was achieved compared to DBP.
SVM has been also studied as another nonlinear
equalizer\cite{Sebald-SVM, Giacoumidis-SVM}, in which a complicated decision
rule like Yin--Yang spiral boundary\cite{Ho-YinYang} can be learned by
kernel-SVM. RBF kernels have been studied in other literature, e.g.,
\cite{Chen-RBF}. HMM-based turbo
cycle-slip recovery\cite{Koike-Slip} offers greater than $2$\,dB gain. 
A stochastic DBP proposed in \cite{Irukulapati-SDBP-2013} exhibits an
outstanding performance by solving inverse NLSE with SSFM, which
adopts MCMC particle representation of
stochastic noise.

\subsection{Modern Deep Learning Applications}

As shown in Fig.~\ref{fig:trend}, there exist a lot of deep learning applications, among which
a limited number of examples are listed below.
DNN was introduced for optical signal-to-noise ratio (OSNR) monitoring in~\cite{Tanimura2016}.
Modulation classification as well as OSNR monitoring was considered in~\cite{Khan2017},
and a deep CNN showed an accurate performance in~\cite{Wang2017}.
Deep learning-based network management and resource allocation were studied in~\cite{Guo2018} and \cite{Yu2018}.
Analogously, traffic optimization based on deep reinforcement learning (DRL) was also considered in~\cite{Luo-RL, Tang-RL}.
Various end-to-end deep learning which jointly optimizes signal constellation and detection have been proposed, e.g., \cite{Ye2018, Karanov2018, Li2018e2e, Jones2018, Chagnon2018},
where denoising auto-encoder (AE) architecture is trained through nonlinear fiber channels.
Also for receiver-end design, many DNN equalizers to compensate for fiber nonlinearity were introduced for coherent or non-coherent optical links, e.g., \cite{RiosMuller, Chuang2017, Kamalov2018, Li2018, Koike-SPPCom18, Koike-ECOC19}.

Note that big data necessary for deep learning are readily available in
high-speed optical communications, where we can obtain gigabits or terabits of 
data in a second\cite{Millar-Tbps}.
In addition, the DNN is massively parallelizable in hardware implementation, which is suited for future optical
communications.
In modern DNN, various techniques have been introduced,
e.g., pre-training, mini-batch, rectified linear unit (ReLU), dropout, batch normalization,
skip connection, inception, adaptive-momentum (Adam) stochastic gradient,
adversarial, and long short-term memory (LSTM) architectures\cite{Hockreiter-LSTM}.


\section{Deep Learning for Nonlinear Compensation}

Similar to the other DNN equalizers, we focus on deep learning for fiber nonlinearity compensation.
This paper has a distinguished contribution over existing literature
as we propose a novel DNN-based TEQ suited for BICM-ID systems where
state-of-the-art LDPC codes are employed.

\subsection{Nonlinear Fiber-Optic Communications System}
\begin{figure}[t]
 \includegraphics[width=\linewidth]{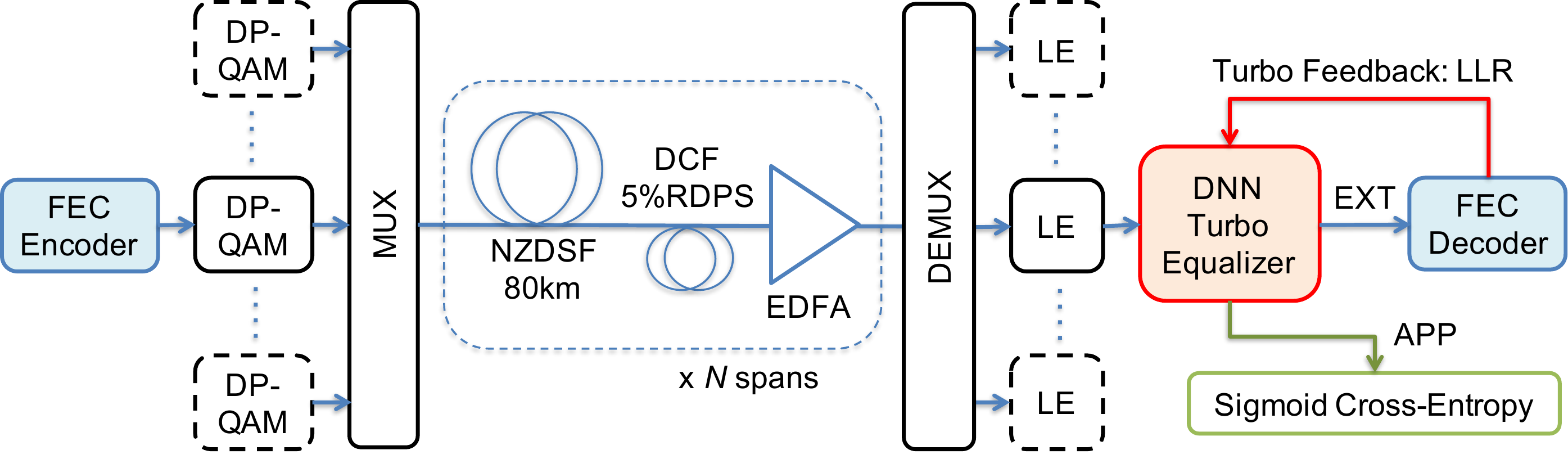}
 \caption{Coherent optical communications with DNN-TEQ.}
 \label{fig:system}
\end{figure}

\begin{figure}[t]
 \centering
 \subfloat[$-5$~dBm Launch]{
 \includegraphics[width=0.3\linewidth]{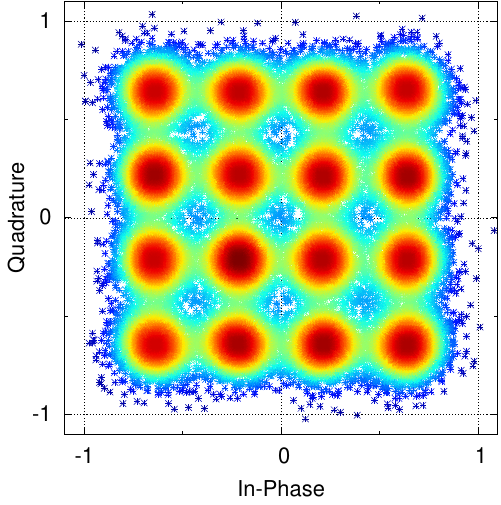}\label{fig:-5}
 }
 \hfill
 \subfloat[$-3$~dBm Launch]{
 \includegraphics[width=0.3\linewidth]{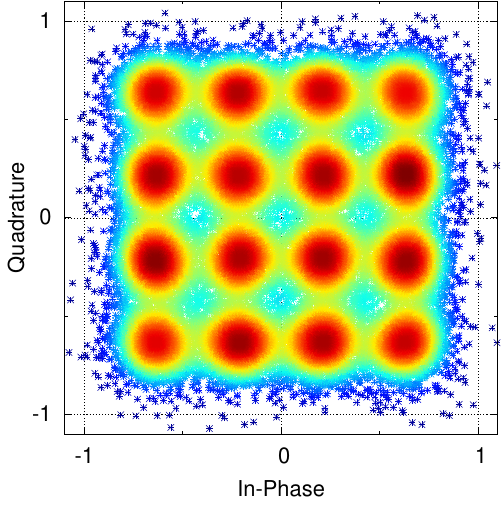}\label{fig:-3}
 }
 \hfill
 \subfloat[$-1$~dBm Launch]{
 \includegraphics[width=0.3\linewidth]{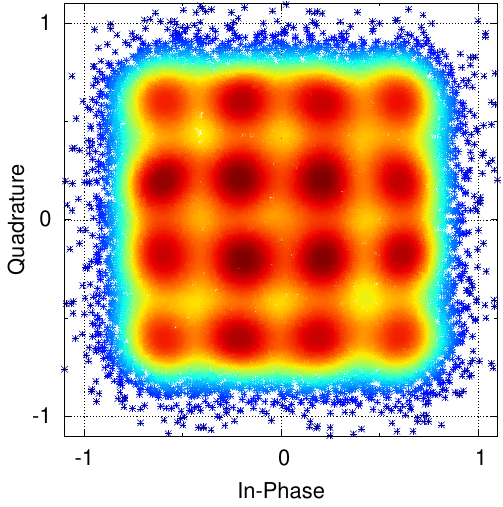}\label{fig:-1}
 }
\caption{Residual distortion of DP-16QAM after LE for $16$-span NZDSF links.}
 \label{fig:const}
\end{figure}

The optical communications system under consideration is depicted in
Fig.~\ref{fig:system}.  Three-channel DP-QAM signals for $34$~GBd baud
rate and $37.4$~GHz channel spacing are sent over fiber plants towards coherent receivers.  We
consider $N$ spans of dispersion managed (DM) links with $80$~km
non-zero dispersion-shifted fiber (NZDSF) at a residual dispersion per
span (RDPS) of $5$\%.
The NZDSF has a dispersion parameter of $D=3.9$~ps/nm/km, a nonlinear
factor of $\gamma=1.6$~/W/km, and an attenuation of $0.2$~dB/km.
The span loss is compensated by Erbium-doped
fiber amplifiers (EDFA) with all ASE noise added just before the receiver assuming the noise figure of $5$\,dB.
We use digital root-raised cosine filters with $10$\% rolloff at both
transmitter and receiver.
The receiver employs standard phase recovery
and linear equalization (LE) to compensate for linear dispersion. Due to fiber nonlinearity, residual distortion
after LE will limit the achievable information rates.

Fig.~\ref{fig:const} shows an example of residual distortion of DP-16QAM
constellation after $31$-tap least-squares LE for $16$-span
transmissions.  We can see that the constellation is more seriously
distorted with the increased launch power due to Kerr fiber
nonlinearity.
%
To compensate for the residual nonlinear distortion, we introduce DNN-based TEQ, which exploits soft-decision feedback from FEC
decoder as shown in Fig.~\ref{fig:system}.

\subsection{Scalable Deep Neural Network Equalization}

\begin{figure}[t]
 \centering
 \subfloat[One $2^{2m}$-ary softmax]{\label{dnn_single}
 \includegraphics[height=0.52\linewidth]{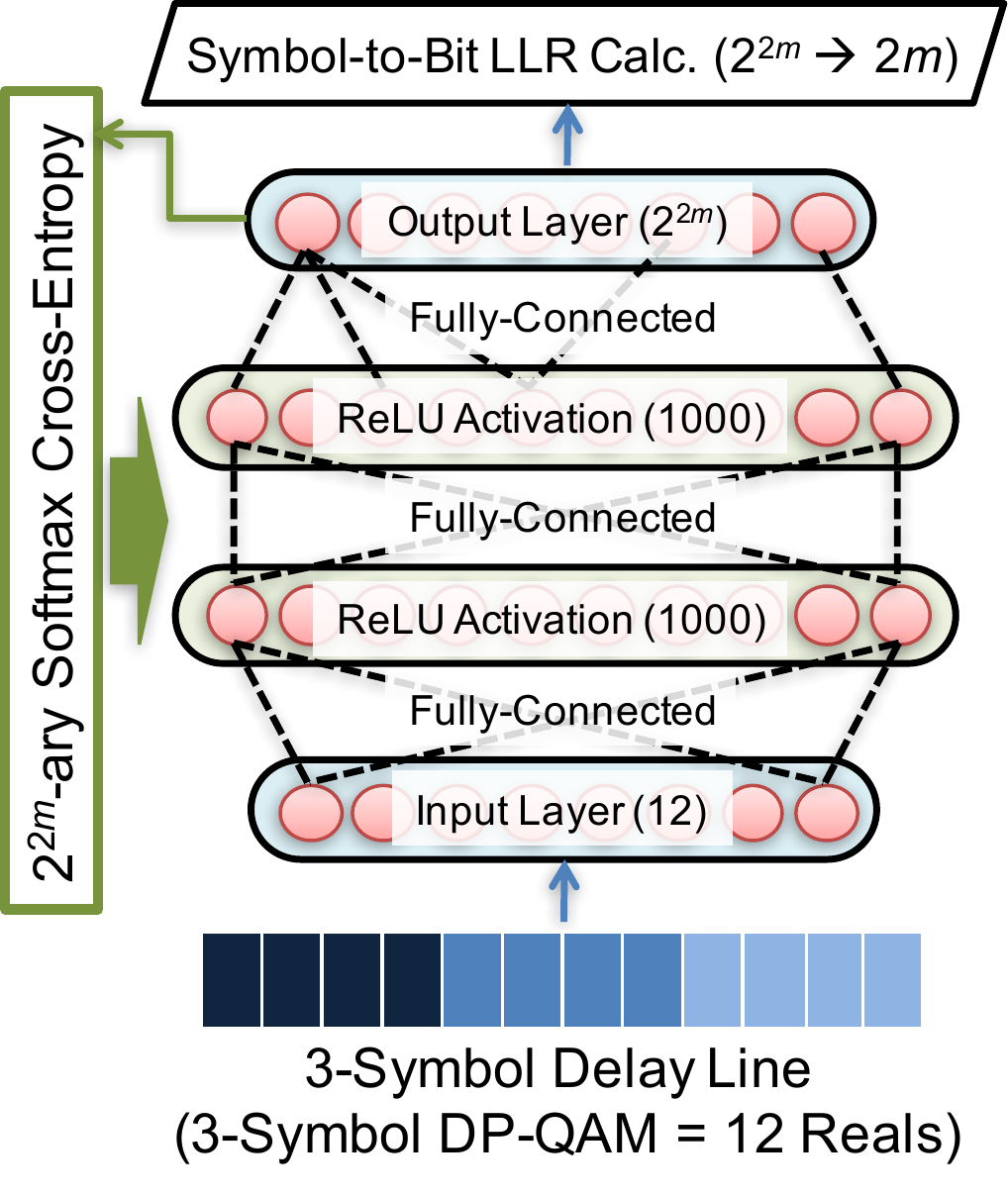}
 }
 \hfill
 \subfloat[$2m \times$ binary softmax]{\label{dnn_multi}
 \includegraphics[height=0.52\linewidth]{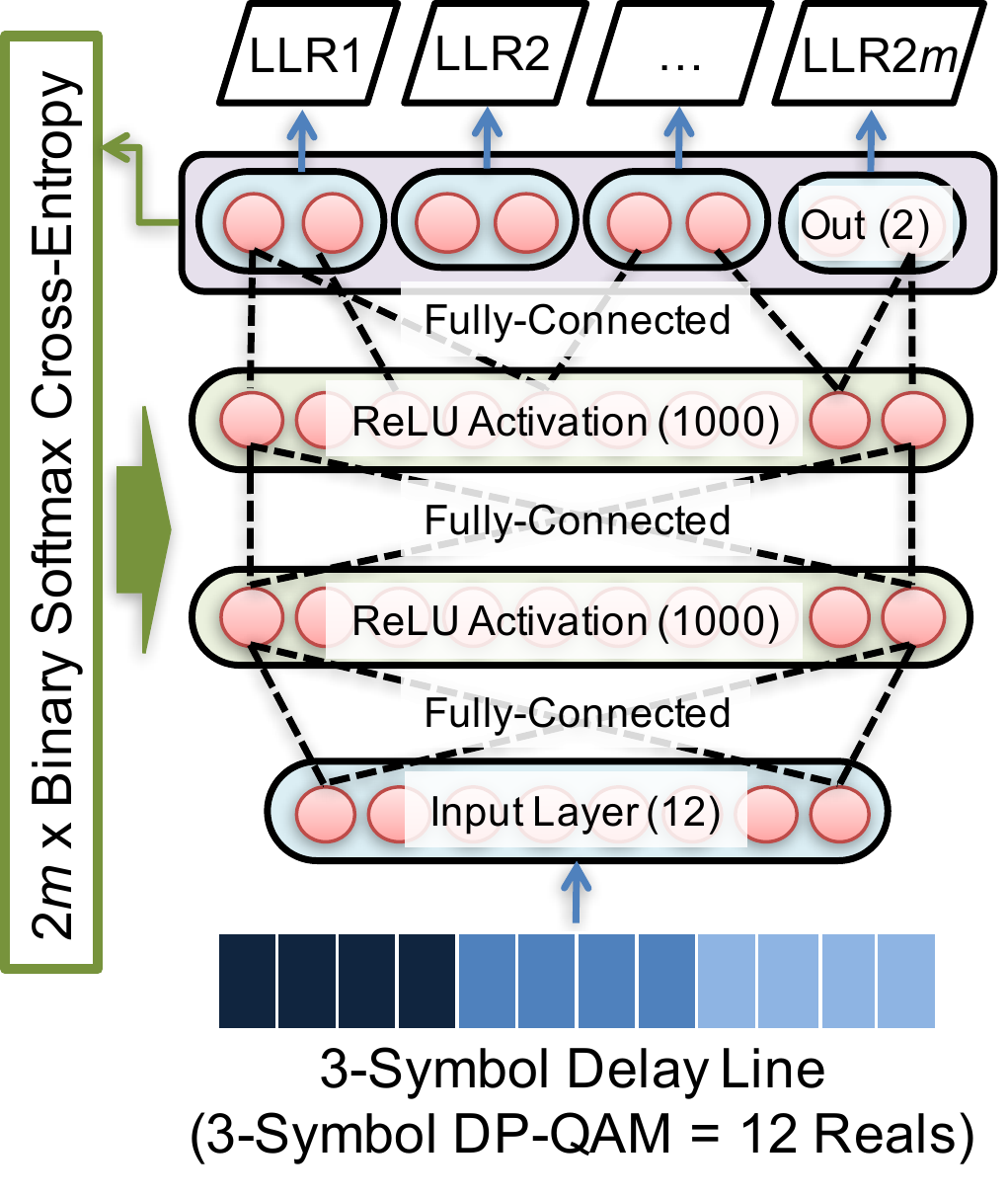}
 }
 \caption{Single-/multi-label DNN for DP-$2^m$QAM.
 }
 \label{fig:dnn}
\end{figure}

Before introducing DNN-TEQ, we discuss loss function to train DNN equalizers suited for BICM.
Consider DP-16QAM equalization, where
there are $8$ bits per symbol, leading to $2^8=256$ classes to identify.
For such multi-class learning, we may use a single nonbinary softmax
classification shown in Fig.~\ref{fig:dnn}(\subref*{dnn_single}), analogous to~\cite{RiosMuller}.
However, this nonbinary (NB) DNN does
not perform well for higher-order DP-QAM in particular for a
limited number of training data.
For example, DP-64QAM requires $4096$ classes to identify
per symbol, which necessitates unrealistically huge data sets for
training.

To be scalable in high-order QAM, we shall use multi-label classification
which employs multiple BCE losses as
shown in Fig.~\ref{fig:dnn}(\subref*{dnn_multi}).
The multi-label DNN produces log-likelihood ratio (LLR),
which can be directly fed into SD-FEC decoder without external processing
such as \cite{RiosMuller, Yoshida}.  This is a great advantage in practice because LLR
calculation is cumbersome, especially for high-order and
high-dimensional modulation.  Note that sum of cross-entropy
minimization is equivalent to maximizing the lower bound of generalized mutual
information (GMI), which is used for SD-FEC performance metric.

\subsection{Nonbinary vs. Binary DNN Equalization}
\begin{figure}[t]
 \centering
  \includegraphics[width=.91\linewidth]{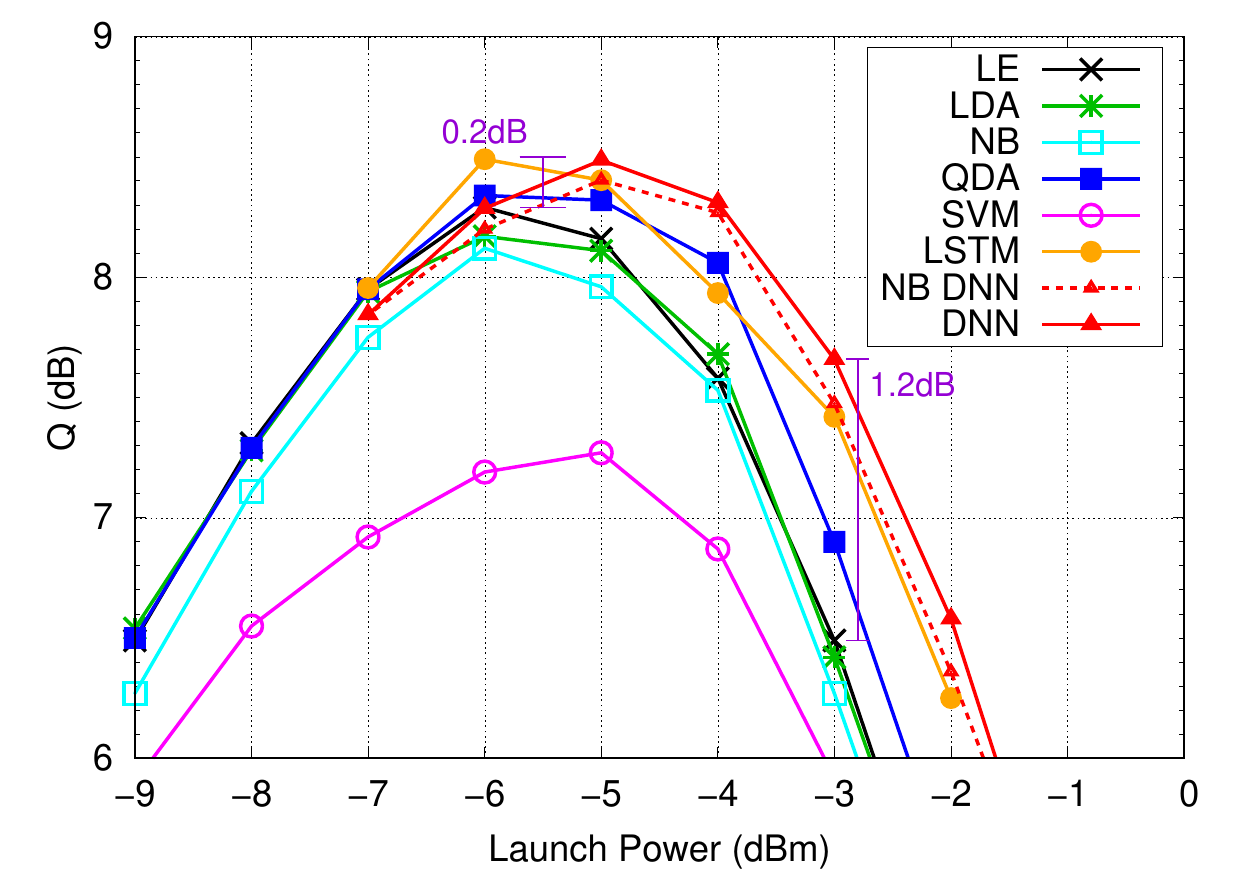}
  \caption{Q factor comparisons for DP-4QAM $50$-span NZDSF.}
  \label{fig:qfac4}
\end{figure}
\begin{figure}[t]
  \centering
  \includegraphics[width=.91\linewidth]{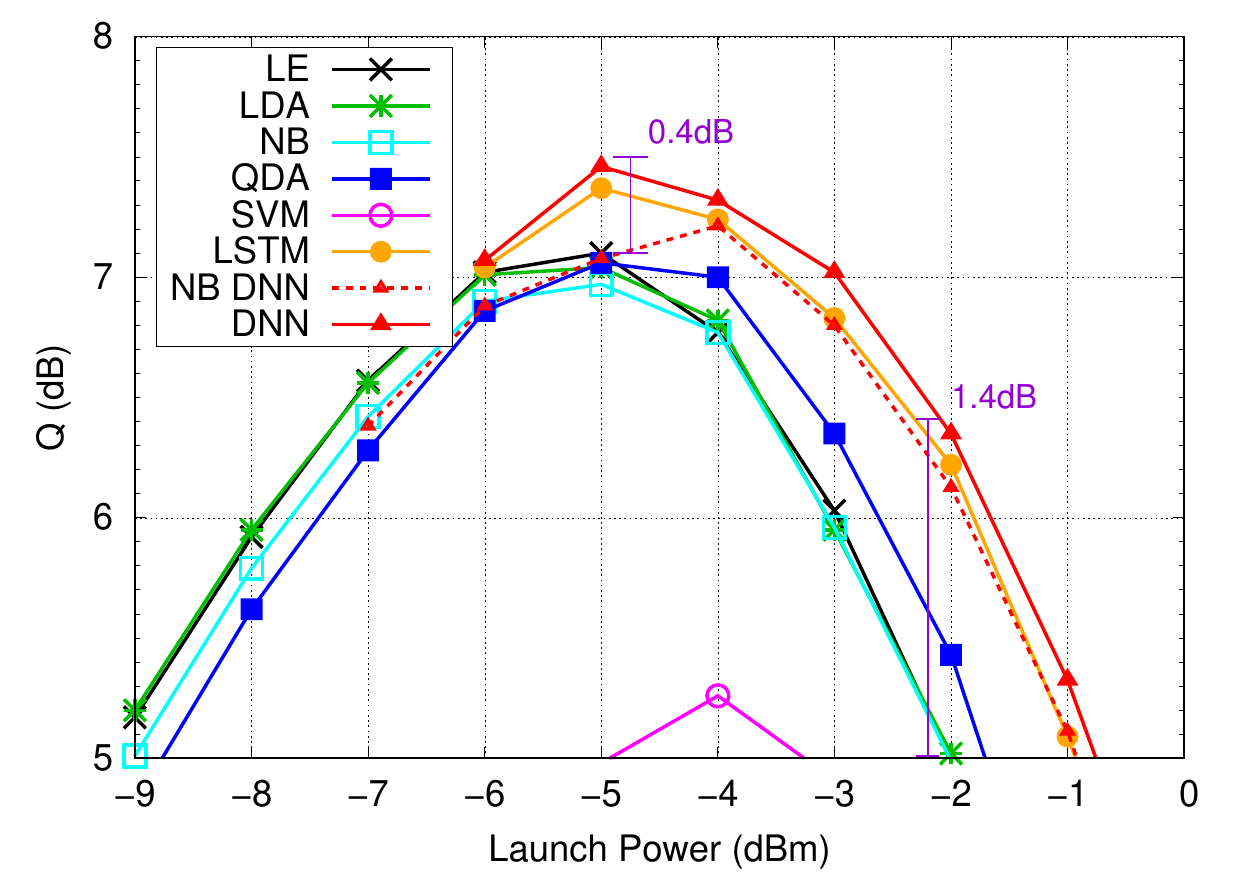}
  \caption{Q factor comparisons for DP-16QAM $16$-span NZDSF.}
  \label{fig:qfac16}
\end{figure}
\begin{figure}[t]
  \centering
 \includegraphics[width=.91\linewidth]{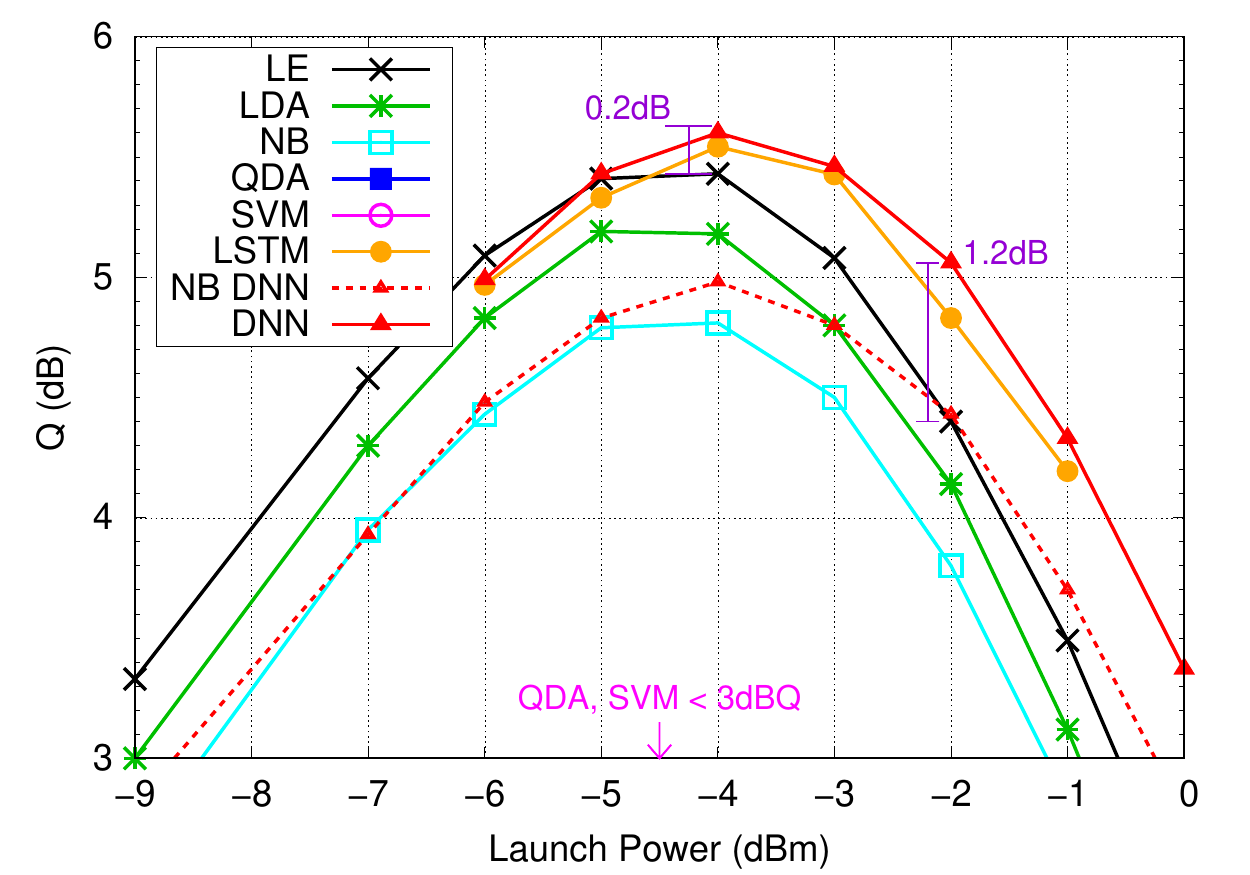}
  \caption{Q factor comparisons for DP-64QAM $8$-span NZDSF.}
 \label{fig:qfac64}
\end{figure}

We compare DNN and LSTM with classical machine learning methods,
specifically, linear discriminant analysis (LDA), na\"{i}ve Bayes (NB),
quadratic discriminant analysis (QDA), and SVM. For multi-class SVM, we
use one-vs-one rule with linear kernel as it worked best among several
variants such as one-vs-all and polynomial kernel. 
The DNN weight is trained by Adam with a dropout ratio of $0.5$ and a batch size of $100$~symbols
to minimize a sum of softmax cross-entropy loss across all labels, using
approximately $5\times 10^5$ training symbols.
Figs.~\ref{fig:qfac4}, \ref{fig:qfac16}, and \ref{fig:qfac64} show the Q
factor versus launch power of DP-4QAM, DP-16QAM, and DP-64QAM,
respectively, for $50$, $16$, and $8$ spans times $80$~km fiber
configurations.  It is observed that DNN can offer the best performance among other methods, achieving greater than $1.2$~dB
gain over LE in highly nonlinear regimes. More importantly,
the conventional DNN with nonbinary softmax cross-entropy does not perform well for high-order QAMs.
It suggests that DNN equalizers using BCE loss function has a great advantage not only for BICM compatibility but also for high-order QAM scalability.

\section{Neural Turbo Equalization: DNN-TEQ}

\begin{figure}[t]
 \centering
 \includegraphics[width=\linewidth]{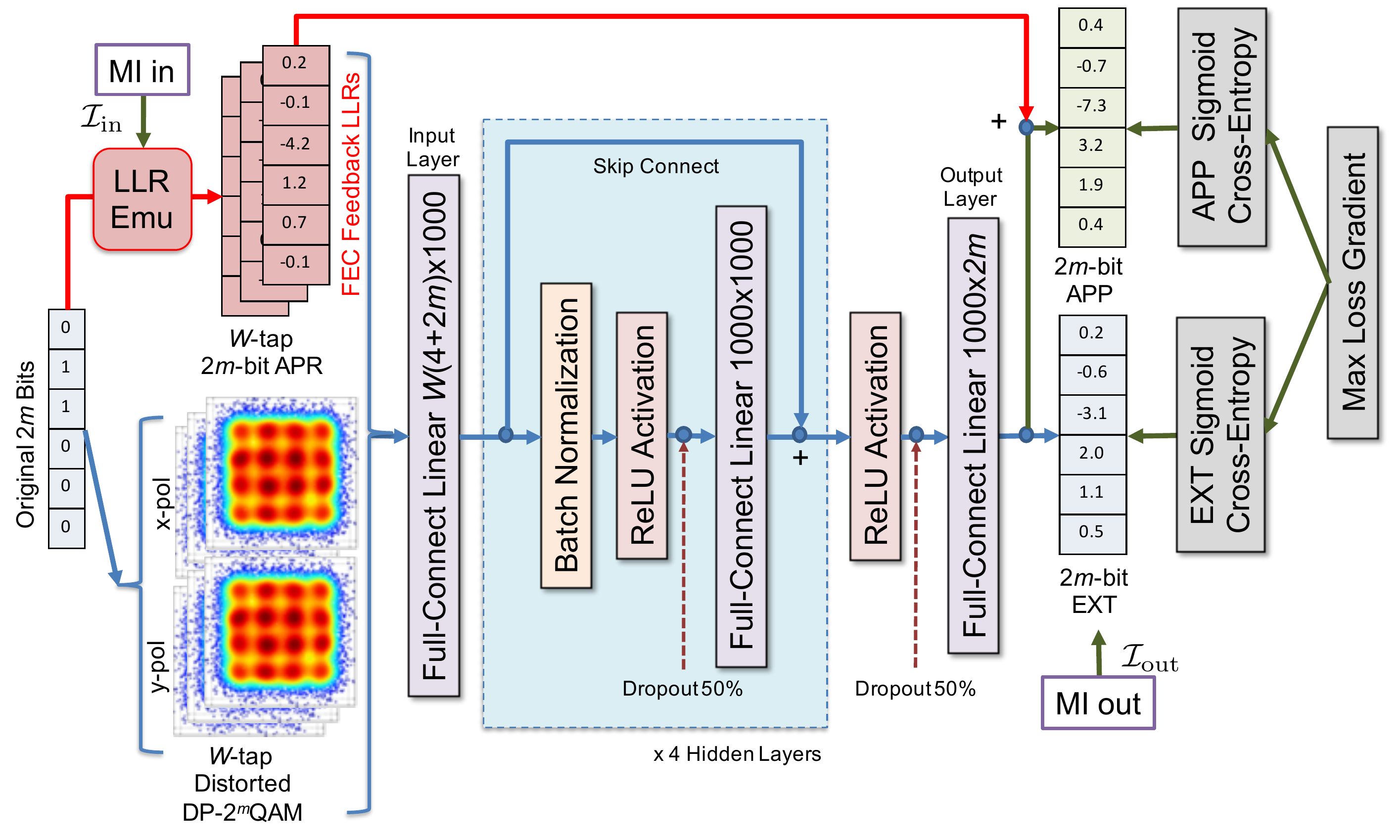}
 \caption{DNN-TEQ architecture and min-max-loss training.}
 \label{fig:teq}
\end{figure}

\subsection{Nested Residual Network Architecture}

Fig.~\ref{fig:teq} shows the architecture of our turbo DNN equalizer, which feeds distorted DP-QAM signals over consecutive $W=3$-tap symbols to generate soft-decision LLR values for FEC decoding.
The major extension from conventional DNN lies in the input layer which takes \textit{a priori} (APR) side information along with DP-QAM symbols.
The APR side information comes from FEC decoder representing intermediate soft-decision LLRs in run time.
For efficient DNN training, the APR values having mutual information of $\mathcal{I}_\mathrm{in}$ are synthetically generated via a Gaussian distribution following $\mathcal{N}((-1)^b\sigma^2/2, \sigma^2)$ where $b$ is an original bit and $\sigma = J^{-1}(\mathcal{I}_\mathrm{in})$ with $J^{-1}(\cdot)$ being ten Brink's J-inverse function~\cite{tenBrink2004}, instead of considering a particular FEC decoder feedback.

The last layer has two branches, i.e., \textit{extrinsic} (EXT) output and \textit{a posteriori} probability (APP) output, which uses a skip connection from the input layer to sum up EXT and APR at a target symbol.
This nested residual network tries to train extrinsic message passing for TEQ realization.
It was found that learning DNN model to minimize APP cross-entropy loss does not always minimize EXT cross-entropy loss accordingly, and vice versa.
In order to keep both APP and EXT outputs reliable, we use a max-pooling layer following sigmoid cross-entropy loss.

The DNN uses four hidden layers, each of which consists of batch normalization, ReLU activation, and a fully-connected linear layer with skip connections and $50$\% dropout for $1000$ neuron nodes.
The DNN is trained with Adam for a mini-batch size of
$1000$~symbols to minimize the worst sigmoid cross-entropy losses
between APP and EXT outputs, using training datasets of approximately $5\times 10^5$
symbols.  An early stopping with a patience of $13$ is carried out up to
a maximum of $500$ epochs.


\subsection{EXIT Chart Analysis}

\begin{figure}[t]
 \centering
 \includegraphics[width=.91\linewidth]{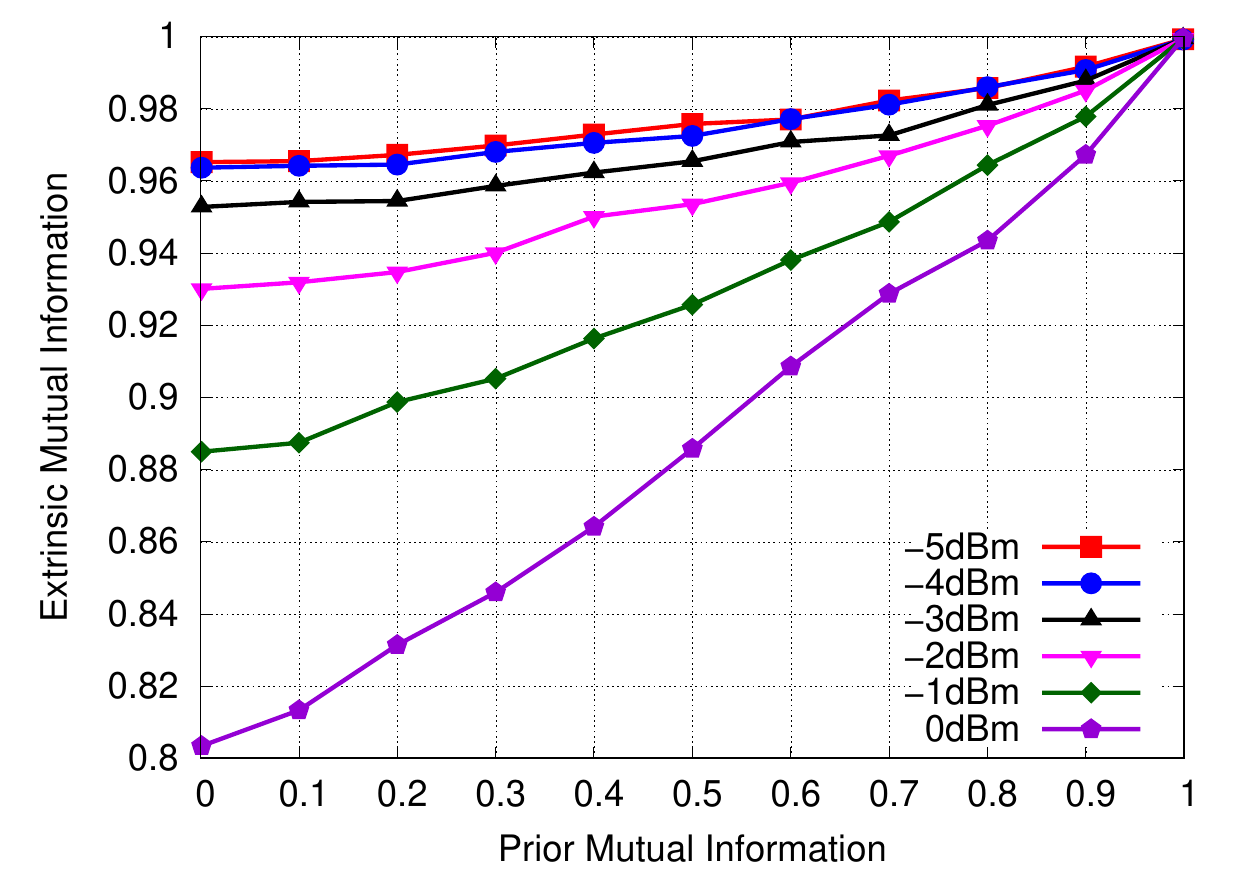}
 \caption{EXIT chart of DNN-TEQ for DP-16QAM in 16-span DM links.}
 \label{fig:exit}
\end{figure}

\begin{figure}[t]
 \centering
 \includegraphics[width=.91\linewidth]{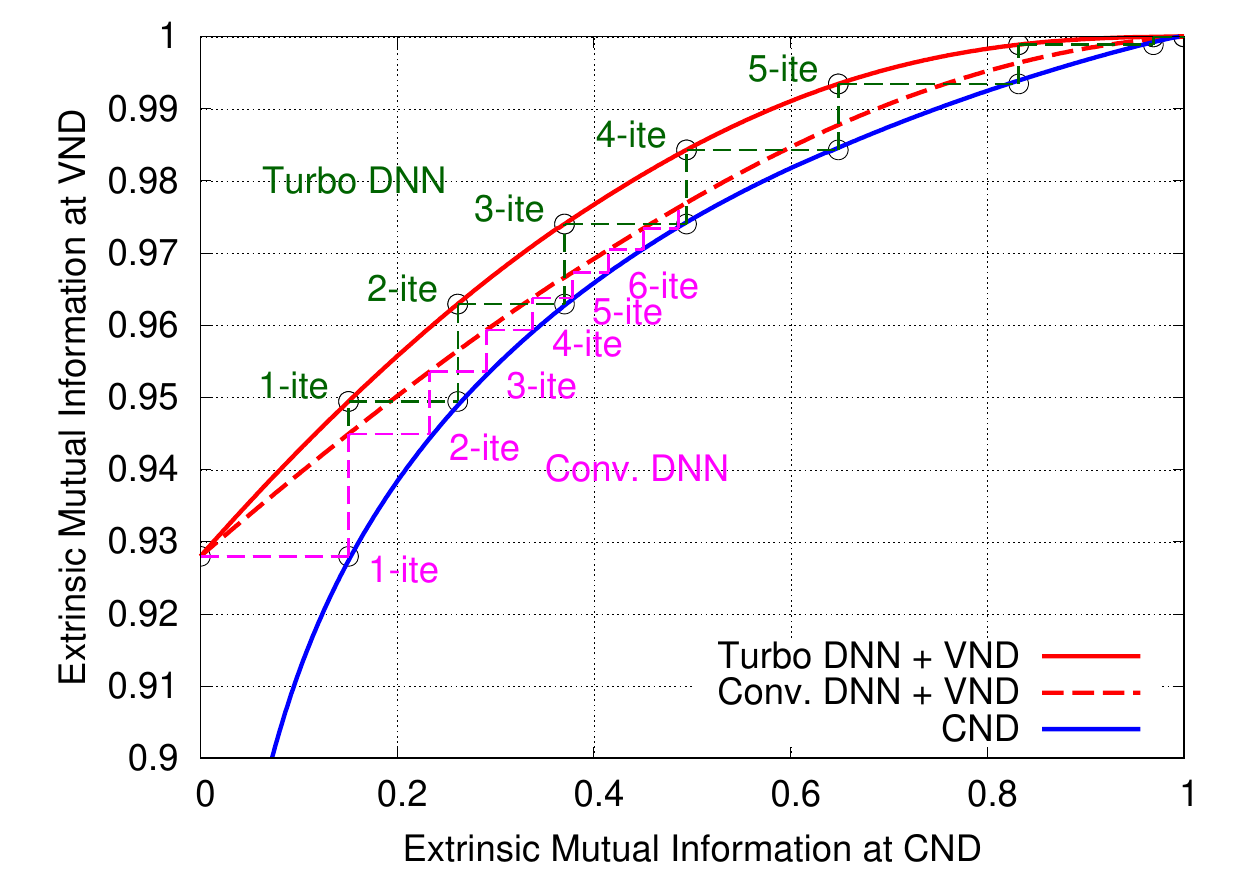}
 \caption{Combined EXIT chart~\cite{tenBrink2004} of DNN-TEQ \& LDPC decoder for DVB-S2 code rate $9/10$ (DP-16QAM in 16-span DM links at $-2$~dBm).}
 \label{fig:traj}
\end{figure}

Fig.~\ref{fig:exit} shows the EXIT chart of DNN-TEQ given LLRs having a certain mutual information from the FEC decoder.
It is clearly observed that the DNN outputs can be greatly improved by feeding in the FEC soft-decision.
An almost linear slope towards $\mathcal{I}_\mathrm{out} = 1$ in EXIT curve is achieved, implying that cross-entropy loss is mitigated linearly with FEC feedback reliability.
This steep slope in the EXIT curve of DNN-TEQ can eventually make a significant improvement in LDPC decoding performance, as shown in Fig.~\ref{fig:traj}, where we present the decoding trajectory between the variable-node decoder (VND) and the check-node decoder (CND) in the LDPC decoder.
Here, we use a combined EXIT chart~\cite{tenBrink2004} of DNN-TEQ and LDPC decoder, for DP-16QAM 16-span DM links at $-2$~dBm launch power and DVB-S2 LDPC codes with a code rate of $9/10$.
As shown, the conventional DNN equalizer without FEC feedback requires a large number of decoder iterations to reach an error-free mutual information of $\mathcal{I}_\mathrm{out} =1$.
Whereas for DNN-TEQ, we can open up an EXIT tunnel between VND and CND curves, that leads to a considerable acceleration of the decoder convergence to reach error-free condition within only a few iterations.

\subsection{BER Performance}

We assume the use of an outer Bose--Chaudhuri--Hocquenghem (BCH) $[30832,30592]$ code with a rate of $0.9922$\cite{Millar-Tbps}, having a minimum Hamming distance of $33$.
Based on the union (upper) bound, the bit-error rate (BER) threshold for this outer BCH code is at or above an
input BER of $5\times 10^{-5}$ to achieve an output BER below $10^{-15}$.
Hence, a post-LDPC BER below $5\times 10^{-5}$ can be successfully decoded
to a BER below $10^{-15}$ when this outer BCH
code is used.

For FEC codes, we consider variable-rate irregular LDPC codes of block length $64{,}800$ bits, used in DVB-S2 standards.
The LDPC codes have a different degree distribution for individual code rates.
For instance at a code rate of $9/10$, the variable degree polynomial (node perspective) is given as $\lambda(x) = 0.1 x^2 + 0.8 x^3 + 0.1 x^4$, whereas the check degree polynomial is $\rho(x)=x^{30}$.
At a code rate of $5/6$, the variable and check degree polynomials are $\lambda(x) = \tfrac{2}{12} x^2 + \tfrac{9}{12} x^3 + \tfrac{1}{12} x^{13}$ and $\rho(x)=x^{22}$, respectively.
We also consider an optimized degree distribution for DNN-TEQ as done analogously in~\cite{tenBrink2004}, where
the EXIT chart of DNN-TEQ in Fig.~\ref{fig:exit} is modeled with cubic functions and EXIT curves of combined VND and DNN-TEQ are optimized for triple-degree check-concentrated distribution, which has two degrees of freedom to search for the best distribution.
For example, the optimized LDPC code for a code rate of $5/6$ at a launch power of $-4$\,dBm for DP-64QAM systems has a degree distribution of $\lambda(x) = 0.725x^2 + 0.25x^9+0.225x^{30}$.

\begin{figure}[t]
 \centering
  \includegraphics[width=.91\linewidth]{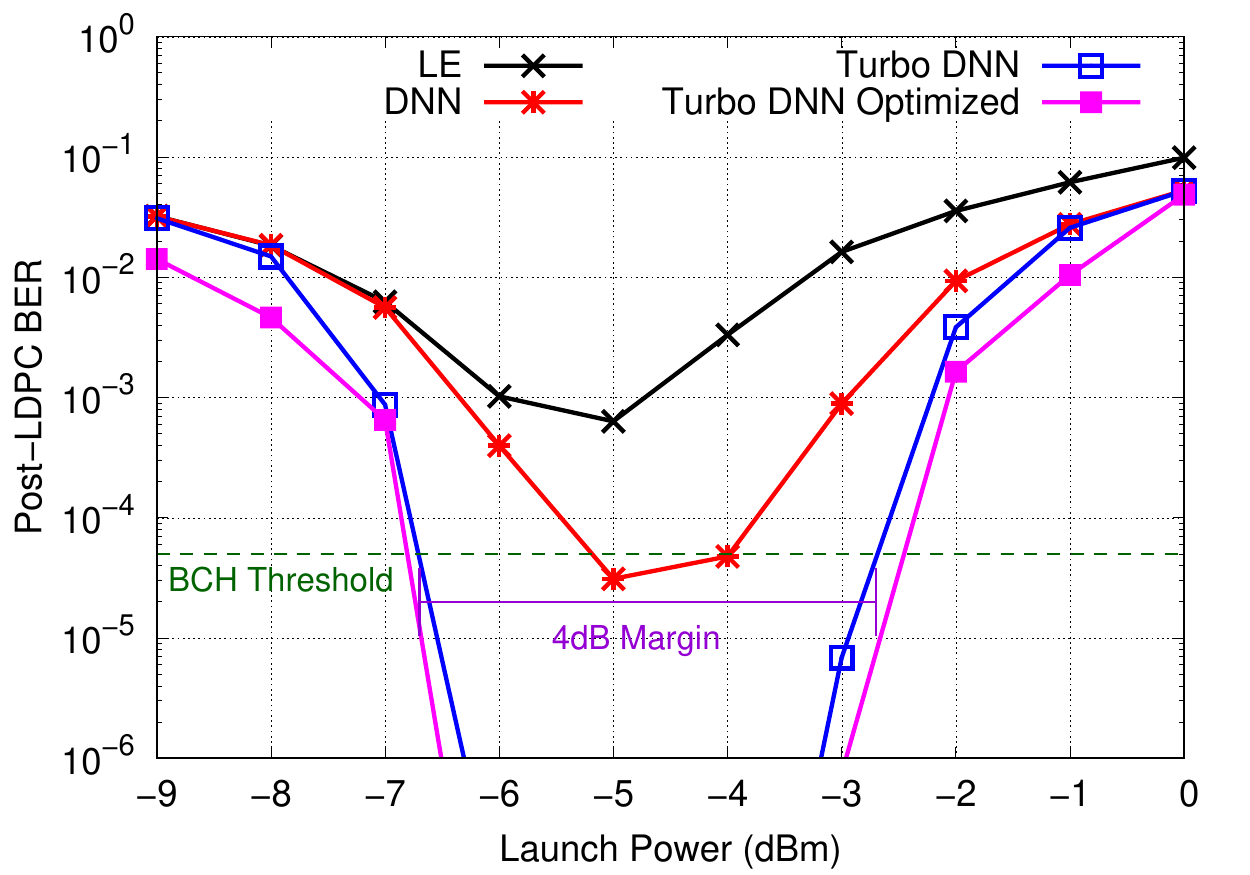}
  \caption{BER performance for DP-16QAM $16$-span NZDSF (LDPC code rate $9/10$, $4$-ite BP).}
  \label{fig:ber16}
\end{figure}
\begin{figure}[t]
 \centering
 \includegraphics[width=.91\linewidth]{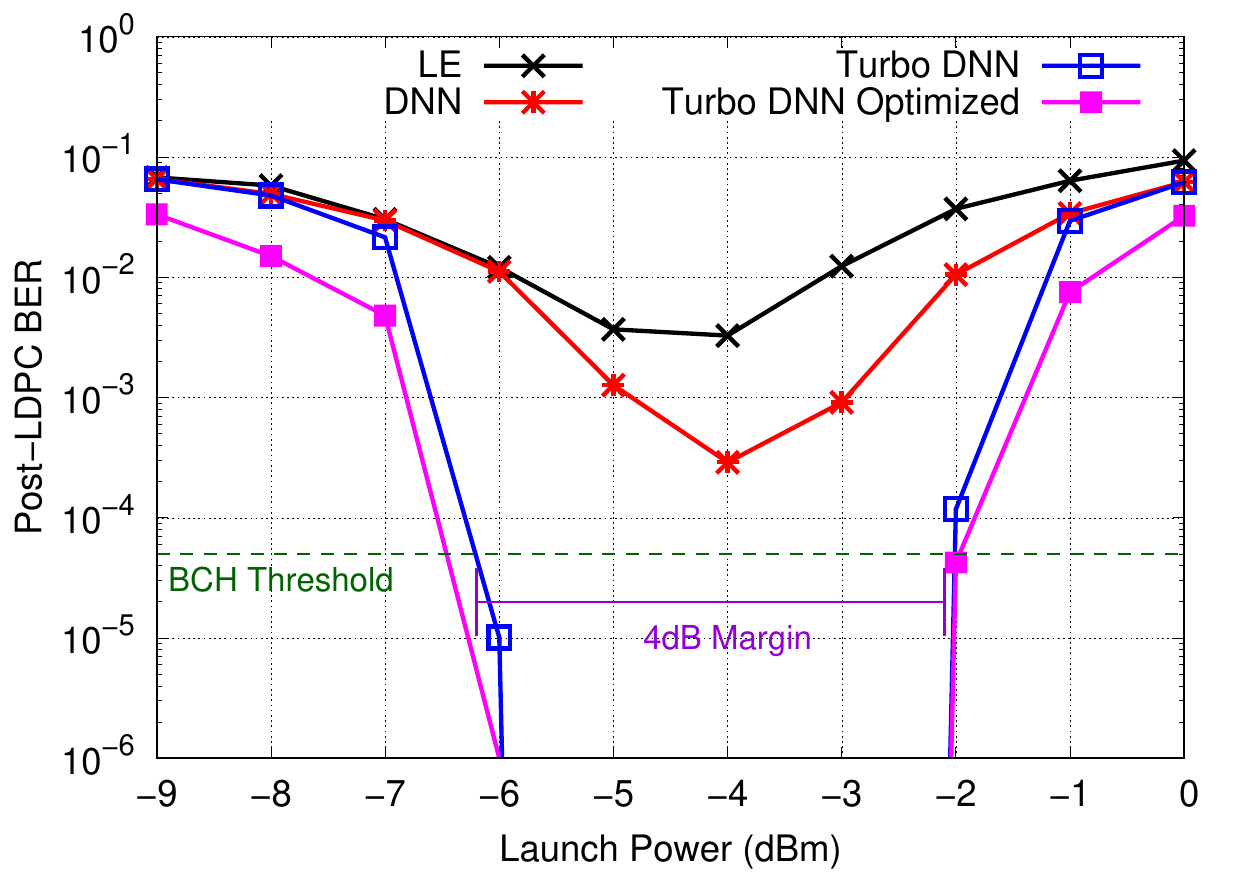}
 \caption{BER performance for DP-64QAM $8$-span NZDSF (LDPC code rate $5/6$, $8$-ite BP).}
 \label{fig:ber64}
\end{figure}

Figs.~\ref{fig:ber16} and \ref{fig:ber64} show the post-LDPC BER
performance versus launch power of DP-16QAM and DP-64QAM,
respectively, for $16$, and $8$ spans of NZDSF links.
We compare DVB-S2 LDPC codes for LE, DNN and DNN-TEQ and our optimized LDPC code for DNN-TEQ.
From the figures, we can observe the following results:
\begin{itemize}
 \item Although DNN nonlinear compensation can improve BER performance of LE, achieving a BER of BCH threshold is mostly in failure.
 \item DNN-TEQ can significantly improve the BER performance of DNN to reach the threshold and about $4\,$dB margin around optimal launch power is realized.
 \item Optimizing LDPC codes for DNN-TEQ can offer an additional marginal improvement over the standard DVB-S2 LDPC codes for the whole range of launch power.
\end{itemize}



%

\subsection{Achievable Rate Performance}

\begin{figure}[t]
 \centering
 \includegraphics[width=.91\linewidth]{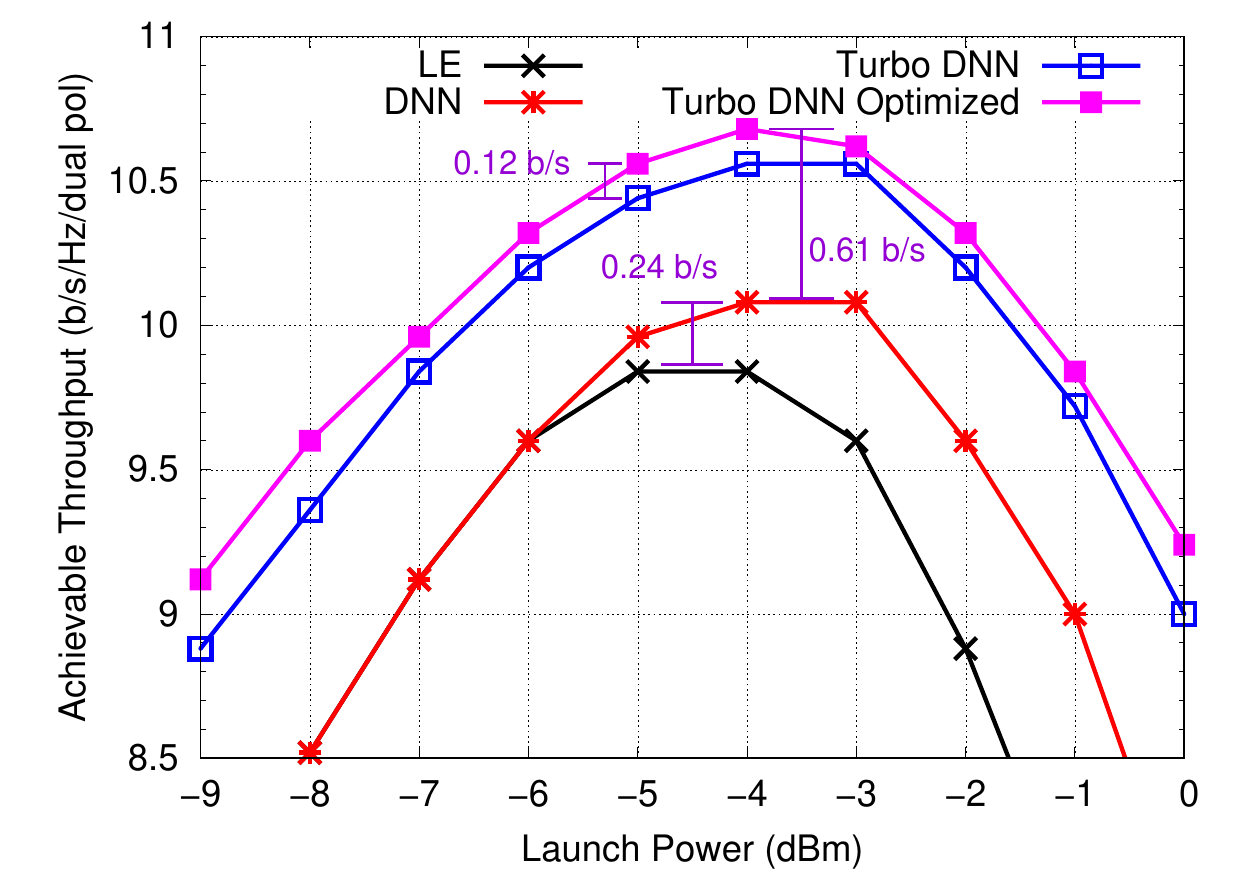}
 \caption{Achievable rate for DP-64QAM $8$-span NZDSF.}
 \label{fig:air64}
\end{figure}

The BER improvement with our proposed DNN-TEQ implies that we can increase the achievable throughput when the code rate is adaptively optimized.
Fig.~\ref{fig:air64} shows achievable rate performance for DP-64QAM at $8$-span NZDSF links.
Here, we use the same variable node degree of DVB-S2 rate $5/6$ and plot the largest code rate such that the post-LDPC BER meets the BCH threshold by varying the check node degree to be a target rate.
From this figure, we can see that the DNN nonlinear compensation can improve the performance of LE by $0.7$\,b/s/Hz in the nonlinear regimes, and the achieved gain in the peak throughput is about $0.24\,$b/s/Hz.
Our DNN-TEQ offers a remarkable BICM-ID gain over the whole range of launch power, achieving a throughput improvement of $0.61\,$b/s/Hz over the DNN when LDPC code is optimized.
A total throughput improvement of $0.85$\,b/s/Hz from the standard LE was achieved by the proposed DNN-TEQ.

\section{Conclusions}

We extended DNN machine learning techniques to TEQ for improved nonlinear compensation in coherent fiber
communications.
We first verified that DNN trained with binary cross-entropy loss can outperform various machine learning techniques to compensate for fiber nonlinearity.
Through EXIT chart analysis, we then confirmed that the proposed DNN-TEQ offers decoder acceleration by feeding intermediate soft-decision LLR from the LDPC decoder.
Our DNN-TEQ significantly improves BER performance through the turbo iteration.
We also investigated LDPC code design to match the EXIT chart of DNN-TEQ, and demonstrated that the proposed DNN-TEQ with optimized LDPC codes can improve the achievable throughput by $0.85$\,b/s/Hz over linear equalization with standard LDPC codes.
To the best of authors' knowledge, this is the first paper investigating TEQ based on DNN for fiber nonlinearity mitigation.

%


\end{document}